\documentclass[twocolumn,prx,showpacs,superscriptaddress,longbibliography]{revtex4-1}

\usepackage{soul}

\usepackage{xcolor}
\colorlet{mylinkcolor}{blue!66!black!80}
\usepackage{mathtools}
\usepackage[colorlinks=true,linkcolor=mylinkcolor,citecolor=mylinkcolor,filecolor=cyan,urlcolor=mylinkcolor,breaklinks=true]{hyperref}
\usepackage[utf8]{inputenc}

\newcommand{\A}{\mathcal{A}}

\newcommand{\kkm}{k_-^{\textrm{eff}}}
\newcommand{\kkp}{k_+^{\textrm{eff}}}

\newcommand{\C}{\mathcal{C}}

\begin{document}
\title{Cost and Precision of Brownian Clocks}
\author{Andre C. Barato}%
\affiliation{%
Max Planck Institute for the Physics of Complex Systems, N\"othnizer Stra{\ss}e 38, 01187 Dresden, Germany}%
 \author{Udo Seifert}%
\affiliation{%
II. Institut f\"ur Theoretische Physik, Universit\"at Stuttgart, 70550 Stuttgart, Germany}

\begin{abstract}

Brownian clocks are biomolecular networks that can count time. 
A paradigmatic example are proteins that go through a cycle thus
regulating some oscillatory behaviour in a living system. Typically,
such a cycle requires free energy often provided by ATP hydrolysis.
We investigate the relation between the precision of such a clock
and its thermodynamic costs. For clocks driven by a constant thermodynamic
force, a given precision requires a minimal cost that diverges as the
uncertainty of the clock vanishes. In marked contrast, we show that
a clock driven by a periodic variation of an external protocol can achieve
arbitrary precision at arbitrarily low cost. This result
constitutes a fundamental difference between processes driven by a
fixed thermodynamic force and those driven periodically. As a main
technical tool, we map a periodically driven system with a deterministic protocol 
to one subject to an external protocol that changes in stochastic time intervals,
which simplifies calculations significantly. In the non-equilibrium 
steady state of the resulting bipartite Markov process, the uncertainty of the clock 
can be deduced from the calculable dispersion of a corresponding current.

\end{abstract}
\maketitle

\section{Introduction}

Periodic behavior is ubiquitous in living systems, from neural oscillations \cite{buzs06} to circadian cycles \cite{gold97,naka05}. 
An example of a well studied biochemical oscillation is the phosphorylation-dephosphorylation 
cycle of the KaiC protein \cite{naka05,dong08,embe06,vanz07,zwic10}. This phosphorylation-dephosphorylation cycle functions as a circadian clock allowing a cyanobacterium 
to tell time \cite{dong08}, i.e., to oscillate in synchrony with day-night changes. Another example of a biochemical oscillation that is related to a 
phosphorylation-dephosphorylation cycle of a protein happens in the activator-inhibitor model recently analyzed in \cite{cao15}. More generally, biochemical 
oscillations are typically associated with a protein that goes through a cyclic sequence of states. Any such protein can be taken as an example of a Brownian clock. 

Brownian clocks are stochastic and, therefore, some uncertainty must be associated with them \cite{potv16}. 
Quite generally, uncertainty related to stochastic changes at the molecular level is an important issue in biophysics. For example,
the Berg and Purcell limit on the maximal precision of a receptor that measures an external ligand concentration  is such a 
fundamental result \cite{berg77,bial05,endr08,endr09,mora10,gove12,kaiz14}. The
relation between precision of some kind and energy dissipation in biophysics has become an active area of research \cite{qian07,cao15,lan12,meht12,bara13b,palo13,lang14,gove14,gove14a,sart14,bara14a,hart15,bo15,ito15}, often using concepts from stochastic thermodynamics \cite{seif12}. 
Specific examples include a relation between energy dissipation and adaptation error in chemotaxis \cite{lan12}, bounds on the precision of estimating an external ligand concentration by a receptor related to energy
consumption \cite{meht12}, a relation between energy dissipation and accuracy in biochemical oscillations \cite{cao15}, and the relation between information-theoretical quantities and 
entropy production in biophysically inspired models \cite{bara14a,sart14,bo15,ito15}. This last example is also related to the growing field of information and thermodynamics
\cite{saga12,mand12,ito13,horo13,deff13,horo14,hart14,bara14,bara14b,rold14,kosk14,parr15}.

The question we investigate in this paper concerns the relation between precision and dissipation in Brownian clocks.
Given that the clock should have a certain precision, what is the minimal energy budget required to run a clock with this precision? 

We model a Brownian clock as an inhomogeneous biased random walk on a ring. The different states of the clock can be interpreted as different 
states of a protein that influences a biochemical oscillation; changes in these states would correspond to, e.g., conformational changes or phosphorylation steps.
We consider two classes of clocks. First, we analyze a clock driven by a constant thermodynamic force that can be generated by, for example, 
ATP. For this class, the general thermodynamic uncertainty relation we obtained in \cite{bara15a} (see also \cite{bara15,ging16,piet16,piet16a,pole16}), establishes
the best precision that can be obtained given a certain energy budget. Within this class a precise clock requires a minimal 
energy dissipation.

The second class is represented by a clock that is driven by a periodic external protocol. Systems driven by such protocols 
reach a periodic steady state and are known as ``stochastic pumps" \cite{parr98,astu07,sini07,sini07a,cher08,raha08,horo09,raha11,astu11,mand14,raz16}. Experimental examples of such systems are 
the generation of rotational motion in an artificial molecular motor driven by an external protocol \cite{leig03} and the pumping of ions across membranes in red blood cells driven by an oscillating electric field \cite{liu90}.
We show that a clock in this class can achieve high precision with an arbitrarily small energy budget. Hence, a clock in this second class is fundamentally different from a clock driven by a fixed thermodynamic force.

The mathematical treatment of systems that reach a periodic steady state, which are driven by deterministic protocols, is typically difficult.
In particular, calculating the dispersion associated with the clock can be quite challenging \cite{zhen13}. For our investigation on the fundamental differences 
between the two classes we consider a generic theoretical framework for which the protocol changes at random time intervals \cite{verl14}. 
Such protocols have been realized in experiments \cite{gome10,diet15}. Within this theoretical framework the system, i.e., the clock, and the external protocol together 
form a bipartite Markov process \cite{bara13b,bara13a,horo14,hart14,horo15}. This property considerably simplifies calculations; in particular, it allows us
to calculate analytically the dispersion of the clock. Using these analytical tools we find the optimal parameters that lead to 
a clock that can achieve high precision with arbitrarily low dissipation. With this proper tuning in hands, we confirm numerically that the 
corresponding clock with a deterministic protocol  can also achieve high precision with vanishing dissipation.

For protocols that change at stochastic times, we prove that given a periodic steady state with 
a certain probability distribution, it is always possible to build a steady state of a bipartite Markov process, which comprises 
the system and  the external protocol, that has the same probability distribution.

This paper is organized as follows. In Sec. \ref{mainsec2} we discuss a clock driven by a fixed thermodynamic force.
Our main result comes in Sec.  \ref{mainsec3}, where we show that a clock driven by an external protocol can combine high precision 
with arbitrarily low dissipation. We conclude in Sec. \ref{mainsec4}. Appendix \ref{App1} contains the thermodynamics of systems driven 
by external stochastic protocols. In Appendix \ref{App2} we prove the equivalence between a periodic steady state and a steady state 
of a bipartite process composed of both system and external protocol. More details for the model analyzed in Sec. \ref{mainsec3} are given 
in Appendix \ref{App3}.

\section{Brownian Clock Driven by a Fixed Thermodynamic Force}
\label{mainsec2}

\begin{figure}
\includegraphics[width=87mm]{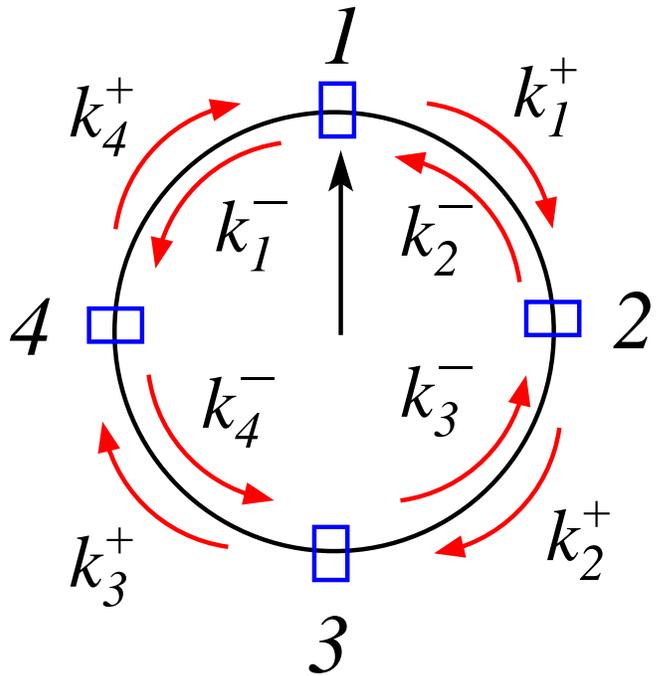}
\caption{Illustration of a Brownian clock with four states. The pointer in state $1$ moves to state $2$ with rate $k_1^+$, or to state $4$ with rate $k_1^-$, and so on.}
\label{fig1main}
\end{figure}

The simplest model of a Brownian clock is a biased random walk on a ring with $N$, possibly different, states and 
arbitrary rates \cite{derr83}, as illustrated in Fig. \ref{fig1main} for $N=4$. The transition rate from 
state $i$ to state $i+1$ is $k_i^+$, whereas the transition rate from $i$ to $i-1$ is $k_i^-$. 
Time is counted by the number of full revolutions of the pointer. Whenever the pointer undergoes the transition from
state $N$ to state $1$, one unit of clock ``time" has passed. Since the clock is stochastic,
a backward step from state $N$ to state $1$ could happen. If, in the next step, the pointer 
moves from $N$ to $1$, one should not attribute the passing of a second time unit to such a
sequence of events. Hence, one counts a backward steps from $N$ to $1$ as a $(-1)$ unit to prevent
such over-counting. The stochastic variable that counts time thus is a fluctuating current $X$ that 
increases by one if there is a transition from $N$ to $1$ and it decreases by one if there is a transition 
from $1$ to $N$.    

In the stationary state, the average $\langle X\rangle$ is given by the probability
current 
\begin{equation}
J\equiv \langle X\rangle/\mathcal{T}= k_N^+P_N-k_1^-P_1, 
\label{eqcurr}
\end{equation}
where the clock runs for a total time $\mathcal{T}$. The inverse $J^{-1}$ is the average time for the clock to complete a cycle, which should 
correspond to the average period of oscillation of the biological 
function regulated by the clock.  An alternative random variable for counting time
would be the cycle completion time which is, however, well-defined only
if  $k_N^-=0$ \cite{foot}.  

Measuring time with this clock comes with a finite uncertainty
\begin{equation}
\epsilon^2 \equiv (\langle X^2\rangle-\langle X\rangle^2)/\langle X\rangle^2= 2D/(J^2 \mathcal{T}),
\label{eq:un}
\end{equation}
where we have introduced the diffusion coefficient  
\begin{equation}
D\equiv (\langle X^2\rangle-\langle X\rangle^2)/(2\mathcal{T}).
\label{eqdiff}
\end{equation}

The clock is driven in the clockwise direction by, for example, a chemical 
potential difference $\A$ that is related to the transition rates by the generalized detailed 
balance condition \cite{seif12}. This condition for this clock reads
\begin{equation}
\A= \ln(\Gamma_+/\Gamma_-),
\label{eq:afff}
\end{equation}
where $\Gamma_\pm= \prod_{i=1}^N k_i^\pm$ and we set Boltzmann constant $k_B$ multiplied by the 
temperature $T$ to $k_BT=1$ in the equations throughout the paper. Each revolution of the clock cost an amount $\A$ of free energy.
Hence running the clock for a total time $\mathcal{T}$ costs an average free energy
\begin{eqnarray}
{\cal C} = J\A \mathcal{T}=\langle X\rangle \A.
\label{eq:cost}
\end{eqnarray} 

The uncertainty of the clock, the cost of running it and its number of states $N$ are constrained
by a universal thermodynamic uncertainty relation \cite{bara15a}, which we discuss in the following.

For a biased random walk with uniform rates $k_+$ and $k_-$, the current is $J=(k_+-k_-)/N$ and 
the diffusion coefficient is $D=(k_++k_-)/(2N^2)$ \cite{derr83}. For this case, the cost ${\cal C}$ in Eq. \eqref{eq:cost} times 
$\epsilon^2$ in Eq. \eqref{eq:un} gives ${\cal C}\epsilon^2= 2D\A/J= (\A/N)\coth[\A/(2N)]$, where we used Eq. \eqref{eq:afff} that implies $\A/N=\ln(k_+/k_-)$. 
It turns out that for a fixed affinity $\A$, this product is indeed minimized for such uniform rates \cite{bara15a}, leading to the uncertainty relation 
\begin{equation}
{\cal C} \epsilon^2\geq (\A/N)\coth[\A/(2N)]\geq {\rm max}(2,\A/N).
\label{ineq}
\end{equation}
We note that this bound is saturated, with ${\cal C} \epsilon^2=2$, for 
a clock close to equilibrium, i.e., in the linear response regime with small $\A$. The implications of Eq. \eqref{ineq} for the design, precision and cost of such a Brownian
clock can best be illustrated by comparing two clocks using familiar notions. Suppose we want to
measure reliably, say with a precision $\epsilon = 10^{-2}$, a time of one hour with either a ``slow" clock that takes one minute for a revolution
or a ``fast" clock that takes only one second. The mean of the stochastic variable $\langle X \rangle$
will be 60 or 3600, respectively. First, the inequality (\ref{ineq}) with (\ref{eq:cost}) implies a structural constraint
on the minimal number of states  $N_{\rm min} =(\epsilon^2\langle X\rangle)^{-1}$ required for a cycle which
turns out to be 167 and 3 for the slow and the fast clocks, respectively. The crucial quantity thus is the
product $N\langle X\rangle$, i.e., the number of elementary steps taken for the measurement. For a 
precision of $10^{-2}$, a clock has to undergo at least $10^4$ elementary steps. A clock counting "minutes"
rather than "seconds" is not necessarily less precise provided its cycle consists of sufficiently many
elementary steps. Second, for a given design, i.e., $N$, the affinity driving the
clock has to be at least 
\begin{equation}
\A_{\rm min}=2 N {\rm arccoth} (\langle X\rangle N \epsilon^2)\geq 2/(\langle X\rangle \epsilon^2) .
\end{equation}
For the slow clock, $\A_{\rm min}\simeq 333$, and for the fast one $\A_{\rm min}\simeq 5.55$. The overall
cost of measuring one hour with this precision is bounded by $20000$ for both types.  From an energetic
point of view, neither the slow nor the fast design is preferable. 

In a biochemical network, free energy is typically provided by $ATP$ hydrolysis, which in physiological conditions liberates approximately $20k_BT$. 
The universal result ${\cal C} \epsilon^2\ge 2$ implies that small uncertainty always has an energetic price associated with it.
An uncertainty $\epsilon$ requires the consumption of $1/(10\epsilon^2)$ ATP molecules. As we show
next, the situation for a clock driven by an external protocol is fundamentally different, since there high precision does not require 
a minimal energy budget.

\section{Brownian Clock Driven by an External Protocol}
\label{mainsec3}

\subsection{Model Definition}

For a Brownian clock driven by an external time-dependent protocol we also consider a ring geometry with $N$ states. The forward transition rates $k_{i,i+1}(t)$ and the backward 
transition rates $k_{i,i-1}(t)$ depend on the time $t$ with a period $\tau$. The energy of site $i$ is denoted $E_i(t)$, whereas the energy barrier between sites $i$ and
$i+1$ is $B_i(t)$. Using the parameters  
\begin{equation}
\epsilon_i(t)\equiv \textrm{e}^{E_i(t)}\qquad\textrm{and}\qquad\chi_i(t)\equiv \textrm{e}^{-B_i(t)}, 
\end{equation}
we fix the rates as
\begin{equation}
k_{i,i+1}(t)=\chi_i(t)\epsilon_{i}(t)\qquad\textrm{and}\qquad k_{i,i-1}(t)=\chi_{i-1}(t)\epsilon_{i}(t). 
\end{equation}

\begin{figure}
\includegraphics[width=87mm]{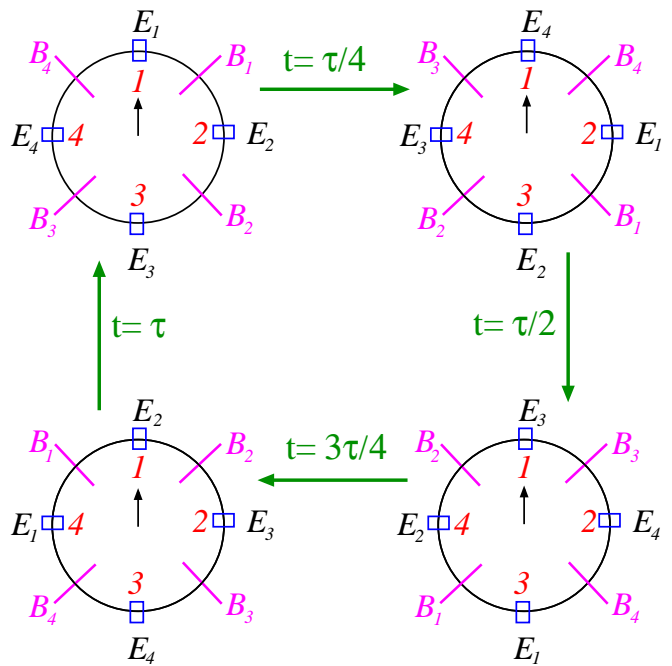}
\caption{Illustration of the deterministic protocol for $N=4$. The red numbers $1,2,3,4$ on the inner cycle denote the position in the clock, 
with the black arrow marking $i=1$ throughout. The energies $E_\alpha$ and the energy barriers $B_\alpha$ rotate one step in the clockwise  
direction after a time interval $\tau/N$, as indicated by the long green arrows. For the variable $\alpha$ these changes represented by 
the long green arrows correspond to an effective backward transition. For example, in the change from the top left to the top right that happens 
at time $t=\tau/4$ the variable $i=1$ remains fixed and the variable $\alpha$ changes from $\alpha=1$ to $\alpha=4$.
}
\label{fig2main}
\end{figure}

For fixed $t$ the rates fulfill detailed balance. Hence, if the rates are time independent, there is no probability current in the ring and 
the clock cannot count time. A current can be generated by a periodic variation of both the energies $E_i$ and the energy barriers $B_i$. 
A simple and symmetric choice for such protocol is as follows, see Fig. \ref{fig2main}. The full period of the external protocol $\tau$ is divided into $N$ parts. 
In the first part of the period from $t=0$ to $t=\tau/N$ the transition rate from state $i$ to state $i+1$ is $k_{i,i+1}(t)=k_i^+\equiv\chi_i\epsilon_{i}$ and the
transition rate from state $i$ to state $i-1$ is $k_{i,i-1}(t)=k_i^-\equiv\chi_{i-1}\epsilon_{i}$. 
In the second part of the period, from $t=\tau/N$ to $t=2\tau/N$ the energies and energy barriers are shifted one step in the clockwise direction, i.e.,
the rates change to $k_{i,i+1}(t)=k^+_{i-1}$ and $k_{i,i-1}(t)=k^{-}_{i-1}$, where for the variable labeling a state $i$ we assume that a sum $i+j$ is modulo $N$. 
In general, the transition rates for $t\in[0,\tau]$ are given by
\begin{equation}
k_{i,i+1}(t)= k^+_{i-j}\qquad\textrm{for}\qquad t\in[j\tau/N,(j+1)\tau/N]  
\end{equation}
and
\begin{equation}
k_{i,i-1}(t)= k^-_{i-j}\qquad\textrm{for}\qquad t\in[j\tau/N,(j+1)\tau/N].  
\end{equation}

Besides the variable $i=1,2,\ldots,N$ we also consider a variable $\alpha=1,2,\ldots,N$, which is convenient for our calculations. Whereas the 
variable $i$ marks a position in the clock the variable $\alpha$ is determined by the energy of the state $E_\alpha$. If the external protocol 
changes during the period, for the variable $i$ the transition rates rotate in the clockwise direction, whereas  the variable $\alpha$ undergoes an 
effective backward transition, as illustrated in Fig. \ref{fig2main}.

The random variable $X$ is the same as for the previous clock: $X$ counts the number of transitions between $i=N$ and $i=1$ in the clockwise direction minus the number of transitions in the anticlockwise 
direction. It turns out that analytical calculations with the above model that reaches a periodic steady state are complicated. In particular, a method to calculate the diffusion coefficient
\eqref{eqdiff} for arbitrary $N$ is not available. 
However, if we consider a protocol that changes at stochastic times with a rate $\gamma=N/\tau$, analytical calculations become simpler. In Appendix \ref{App1}, we 
explain a general theory for such stochastic protocols, along the lines of \cite{verl14}. We show that an analytical expression for the diffusion constant $D$ can be obtained in this case. 
Furthermore, in Appendix \ref{App2} we show that given a periodic steady state arising from a continuous deterministic periodic protocol,
it is always possible to build a bipartite process comprising the system and the stochastic protocol that has the same probability distribution 
as the periodic steady state \cite{foot2}.

\begin{figure}
\includegraphics[width=87mm]{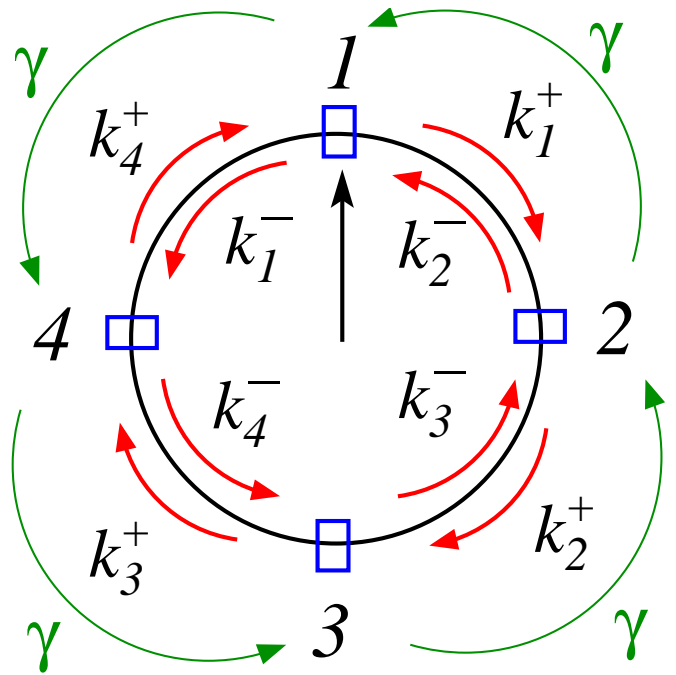}
\caption{Effective network for a clock driven by an external protocol that changes at stochastic times with $N=4$ states.
The green backward arrows represent a jump with rate $\gamma=N/\tau$. A backward jump is equivalent to a 
forward rotation of the rates represented in Fig. \ref{fig2main}. 
}
\label{fig3main}
\end{figure}

For the clock with stochastic protocol, the energies and energy barriers change at stochastic times, with a rate $\gamma=N/\tau$. 
The precise definition of the model for general $N$ is presented in Appendix \ref{App3}. Here in the main text we discuss the case 
$N=4$ that is represented in Fig \ref{fig3main}. It turns out that the full bipartite process can be reduced to a Markov process with four states only.
In this reduced description we use the variable $\alpha$.  The transition rates $\gamma$ are related to one rotation of the transition rates. Effectively, 
such a rotation corresponds to a backward jump of this $\alpha$ variable,  as illustrated for the deterministic protocol in Fig. \ref{fig2main} and explained 
in more detail in Appendix \ref{App3}.

\subsection{Optimal Time-Scales and Energy Barriers}

As explained in  Appendix \ref{App3}, we can calculate current $J$,  entropy production rate $\sigma$, and diffusion constant $D$ analytically for this 
clock with the stochastic protocol, which lead to the product $\C\epsilon^2=2D\sigma/J^2$ as a function of the transition rates. The entropy production is equal to the rate of work done on the system due to 
the periodic variation of the external protocol. Similar to the previous clock driven by a fixed thermodynamic force, if this clock runs for a time $\mathcal{T}$, the energetic cost 
is $\C=\sigma \mathcal{T}$ and the uncertainty is $\epsilon^2=2D/(J^2\mathcal{T})$.

For the simplest clock with  $N=3$, the minimum value of the product turns out to be $\C\epsilon^2\simeq 1.33651$, which is smaller than the universal limit $2$ for systems driven by a fixed thermodynamic force.
We have obtained this product  as a function of the transition rates up to $N=6$. Minimizing $\C\epsilon^2$ numerically, 
we find that the minimum decreases with $N$, and that the transition rates at the minimum have the properties
$\chi_1=\chi_2=\ldots=\chi_{N-1}=\chi\gg \gamma$ and $(\chi_N)^{-1}\to 0$. Thus, in this limit, the energy barrier between states $N$ and $1$ becomes infinite, effectively blocking
transitions between these states. Moreover, the internal transitions are much faster than changes in the protocol, i.e., the system equilibrates before the next change in 
the external protocol happens, which is common in studies about periodically driven systems \cite{parr98,astu07,sini07,sini07a}. 
For this clock, the product $\C\epsilon^2$ is minimized in the far from equilibrium regime, in contrast to the clock from Sec. \ref{mainsec2}, for which the minimum occurs in the linear response regime.

In this limit, the expressions for current $J$ and diffusion coefficient $D$ become
\begin{equation}
J=\gamma Z^{-1}N^{-1}\left(\sum_{\alpha=2}^{N}e^{-E_\alpha}-(N-1)e^{-E_1}\right), 
\label{Jlim}
\end{equation}
and
\begin{equation}
D=\frac{1}{2}\gamma Z^{-1}N^{-2}\left(\sum_{\alpha=2}^{N}e^{-E_\alpha}+(N-1)^2e^{-E_1}\right),
\label{Dlim}
\end{equation}
where $Z\equiv\sum_{\alpha=1}^{N}e^{-E_\alpha}$. These expressions can be obtained by mapping the model in this special limit onto a
biased random walk, as explained in Appendix \ref{App3}. The basic idea behind this mapping is to consider the position of the particle, i.e., the state of the clock, in relation to the barrier. 
If the barrier moves and the particle is in state $\alpha=1$, then the particle crosses the barrier and moves to state $\alpha=N$, corresponding to a backward step of size $N-1$ of the random walk. 
Otherwise, the particle moves one step closer to the barrier, i.e., from state $\alpha$ to $\alpha-1$, corresponding to a forward step of size $1$. 

The entropy production $\sigma$ is calculated with the expression in Eq. \eqref{sgen2}, which gives
\begin{equation}
\sigma=\gamma Z^{-1}\left(\sum_{\alpha=1}^{N}e^{-E_{\alpha}}(E_{\alpha-1}-E_{\alpha})\right). 
\label{Slim}
\end{equation}
This expression for the entropy production, which is the rate of work done 
on the system, can be understood as follows. If there is a jump that changes the external protocol, the work done on the system is given by the 
energy change of the system after the jump. If the system is in a state $\alpha$, this energy change is $E_{\alpha-1}-E_{\alpha}$. Therefore, the rate of work done 
on the system in Eq. \eqref{Slim} is $\gamma$ times a sum over all state $\alpha$ of this energy difference multiplied by the probability of the system 
being in state $\alpha$ before an external jump, which is $Z^{-1}e^{-E_\alpha}$. In marked contrast to the clock driven by a fixed thermodynamic force, the cost $\C=\sigma \mathcal{T}$ for 
this periodically driven clock is, in general, not proportional to the current $J$ that is given in Eq. \eqref{Jlim}.

\subsection{Dissipation-less Clock I: Simple Profile}

Before discussing the optimal energy profile that minimizes the product  $\C\epsilon^2$ we consider the simple profile 
\begin{equation}
E_\alpha= E \delta_{\alpha,1},
\label{eqprofile}
\end{equation}
where $\delta_{\alpha,1}$ is the Kronecker delta. In this case, using Eqs. \eqref{Jlim}, \eqref{Dlim}, and \eqref{Slim} the product $\C\epsilon^2=2D\sigma/J^2$ becomes
\begin{equation}
\C\epsilon^2= \frac{[1+e^{-E}(N-1)]E}{(N-1)(1-e^{-E})}.
\label{prodfor1}
\end{equation}
This expression implies a fundamental difference between the two kinds of clocks. If we choose the parameters $E$ and $N$ in such a way that $e^{E}\gg N\gg E$, 
the product \eqref{prodfor1} can reach an arbitrarily small value. For example, for $N=64$ and $E=5.7$ we obtain $\C\epsilon^2\simeq 0.11$. 
The fact that it is possible to build a clock that has small uncertainty and dissipates arbitrarily low energy is
the main result of this paper. Such a dissipation-less clock is in stark contrast with a clock driven by a fixed 
thermodynamic force, which is constrained by the thermodynamic uncertainty relation $\C\epsilon^2\ge 2$. 

A physical explanation for this result is as follows. Let us consider the case where $E$ is large enough so that the particle is practically never at position 
$\alpha=1$ when the barrier moves forward. This condition amounts to $e^{E}\gg N$. In this case, the position of the particle with respect to 
the energy barrier always diminishes by one when the barrier moves. The current is then given by the velocity of the barrier $J\simeq\gamma/N$ and the 
dispersion is $D\simeq\gamma/(2N^2)$, which is the dispersion of the random walk performed by the barrier that has only forward transitions with rate $\gamma$. 
Work is done on the system only 
if the particle is at state $\alpha=2$ when the barrier moves, which happens with probability $1/(N-1)$. For large $N$, the entropy production is then given 
by $\sigma\simeq\gamma E/N$. The product of cost and uncertainty becomes $\C\epsilon^2= 2D\sigma/J^2\simeq E/N$. The condition $N\gg E$ guarantees a small 
dissipation, leading to a product $\C\epsilon^2$ that can be arbitrarily close to $0$. The mechanism that allows for this scaling of the product  
$\C\epsilon^2$ with $N$ is the large energy barrier that determines the current $J$ and the dispersion $D$. Such a mechanism cannot be realized with the 
clock driven by a fixed thermodynamic force from Sec. \ref{mainsec2}.


\subsection{Dissipation-less Clock II: Optimal Profile}

\begin{figure}
\includegraphics[width=87mm]{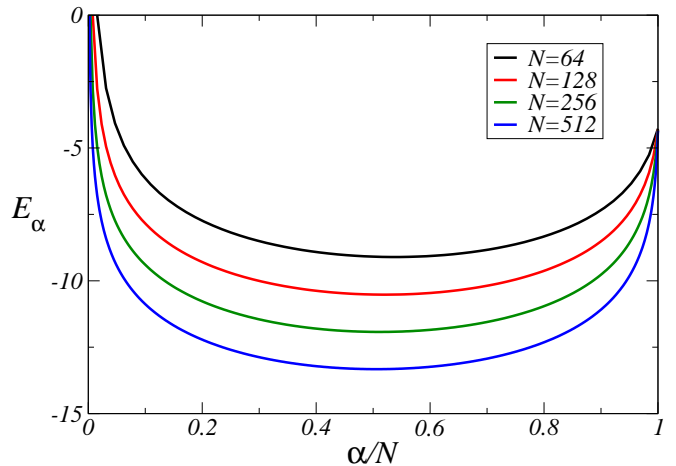}
\caption{Optimal profile $\{E_\alpha\}$ for which the product $\C\epsilon^2$ is minimized.}
\label{fig4main}
\end{figure}

In the limit where the expressions \eqref{Jlim}, \eqref{Dlim}, and \eqref{Slim} are valid, the minimum of $\C\epsilon^2$ is achieved with an optimal energy 
profile $\{E_\alpha\}$ that depends on $N$, as shown in Fig \ref{fig4main}. The negative value of the minimum of this energy profile grows with $N^{2}$, and for larger 
$N$ the profile becomes flatter in the middle. Hence, for large $N$, the probability $P_1$ to be in the state with highest energy  
goes to zero and, from expressions \eqref{Jlim} and \eqref{Dlim}, $J\to \gamma/N$ and $D\to\gamma/(2N^2)$, respectively. Current and diffusion 
are then determined by the unidirectional random walk performed by the barrier, as is the case of the simple profile from Eq. \eqref{eqprofile} with a large $E$.  

We verified numerically that for this optimal profile the entropy production rate behaves as $\sigma\sim N^{-2}$. The product $\C \epsilon^2=2D\sigma/J^2\sim N^{-2}$ can then become 
arbitrarily small for large $N$. For example, for a clock with $N=64$ states and with an optimal energy profile, we get $\C\epsilon^2\simeq 0.0047$. Hence,
with this clock, an uncertainty $\epsilon=10^{-2}$ costs approximately $47k_BT$, which is much less then the minimal 
cost of $20000k_BT$ found above for a clock with the same precision and driven by a fixed thermodynamic force.

This clock with an optimal energy profile also relies on the mechanism of a large barrier that controls the dispersion and current of the clock, with the
difference that the energy dissipation can be suppressed as $N^{-2}$. There are other energy profiles that lead to a dissipation-less and precise clock. 
If we choose a simple energy profile $E_\alpha= -\alpha/N^\phi$, with $0<\phi<1$, from Eqs. \eqref{Jlim}, \eqref{Dlim}, and \eqref{Slim}, 
we obtain $\C \epsilon^2\sim N^{-\phi}$. 

A dissipation-less and precise clock can also be obtained with a deterministic protocol. We have confirmed with numerical simulations up to $N=8$, using the optimal 
energy profile from Fig. \ref{fig4main}, that for a deterministic protocol $J$ and $\sigma$ are the same as given by \eqref{Jlim} and \eqref{Slim}, while $D$ 
becomes smaller. Such a smaller diffusion comes from the fact that the deterministic protocol does not have the randomness associated with 
the waiting times for a change in the protocol. Therefore, the product $\C \epsilon^2$ is even smaller in this case 
and also vanishes for large $N$.

\subsection{Numerical Case Study}

\begin{figure}
\includegraphics[width=87mm]{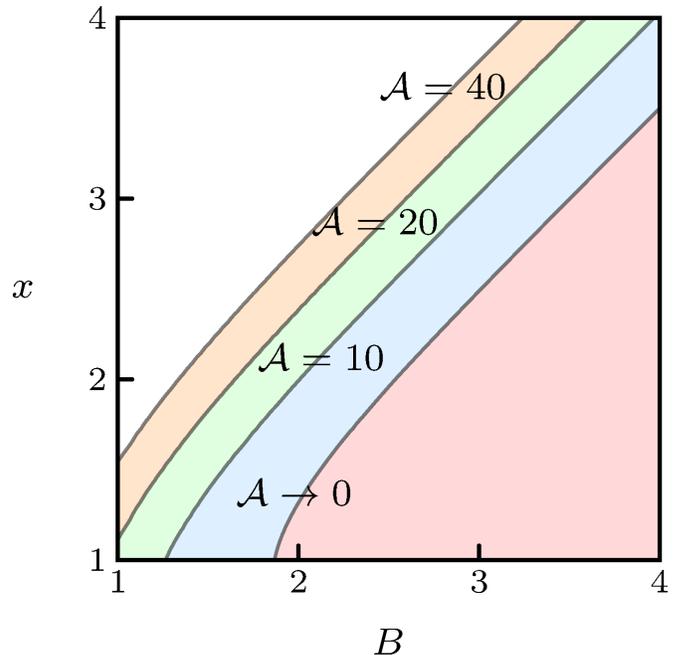}
\caption{Product $\C\epsilon^2$ for a clock driven by an external protocol. 
The parameters of the clock are set to $\chi_1=\chi_2=1$, $\chi_3=10^{-B}$, $\gamma=10^{-x}$, $E_1=0$, $E_2=-1.21938$, and $E_3=-1.43550$.
Below the lines, the product $\C\epsilon^2$ is smaller than
$(\A/3)\coth(\A/6)$, which is the optimal value of this product for a clock driven by a fixed affinity $\A$ and $N=3$.
}
\label{fig5main}
\end{figure}

For illustrative purposes we compare a specific clock driven by an external protocol with the results for clocks driven by a fixed thermodynamic force. 
In Fig. \ref{fig5main}, we show a contour plot of the product $\C \epsilon^2$ for $N=3$. The energies of the clock are set to $E_1=0$, $E_2=-1.21938$,
and $E_3=-1.43550$, which is the optimal profile for $N=3$. The parameters $B$ and $x$ determine the other transition rates in the following way.
The parameters related to the energy barriers are set to $\chi_1=\chi_2=1$ and $\chi_3=10^{-B}$. The rate of change of the protocol is 
set to $\gamma=10^{-x}$. Hence, for large $B$ and $x$, the product  $\C \epsilon^2$ reaches its minimal value 
for $N=3$, which is $\C \epsilon^2\simeq1.33651$.

This externally driven clock can be compared to an optimal clock driven by a fixed thermodynamic force $\A$ with the same number of states $N=3$.
The product $\C\epsilon^2$ for the optimal clock driven by a fixed affinity $\A$ saturates the inequality \eqref{ineq}, i.e., 
for $N=3$ this optimal clock follows the relation $\C\epsilon^2=(\A/3)\coth(\A/6)$, which is an increasing function of the affinity. Close to equilibrium, $\A\to 0$, the product reaches the minimal value 
$\C\epsilon^2=2$.  Hence, a clock driven by a fixed thermodynamic force cannot have a better tradeoff relation between cost and precision than the externally driven clock  inside the region limited by the line 
$\A\to 0$ in Fig. \ref{fig5main}. Increasing the affinity $\A$ leads to a larger region  for which the externally driven clock 
has an smaller product $\C\epsilon^2$.

\section{Discussion and Conclusion}
\label{mainsec4}

We have shown that a Brownian clock driven by an external protocol can achieve small uncertainty in a dissipation-less manner. This result 
constitutes a fundamental difference between systems driven by a fixed thermodynamic force and systems driven by an external protocol. For the first 
case, small uncertainty does have a fundamental cost associated with it, which is determined by the thermodynamic uncertainty relation from \cite{bara15a}.

More realistic models related to biochemical oscillations do not typically have a simple space of states like 
the ring geometry considered in this paper. However, this feature does not represent a limitation in our fundamental bounds. 
First, the thermodynamic uncertainty relation $\C\epsilon^2\ge 2$ is not limited to the ring geometry but valid 
even for any multicyclic networks of states \cite{bara15a,ging16}. Second, we have shown that it is possible to reach $\C\epsilon^2\to 0$ with a specific model, which is 
sufficient to prove that systems driven by an external periodic protocol can, in principle, achieve high precision 
with vanishingly small dissipation.

Main features of the protocol that achieves high precision in a dissipation-less manner are internal transitions much faster 
than changes in the external protocol, a large number of states, and a large energy barrier that effectively blocks transitions between one pair of states. This 
third property does not allow for cycle completions without a change in the external protocol. It remains to be seen whether further
classes of protocols that also lead to $\C\epsilon^2\to 0$ exists. In particular, a quite different externally driven system, known as a hidden pump, that leads
to a finite current with an arbitrarily low entropy production has been proposed in \cite{espo15}. It would be worthwhile to verify whether such hidden pumps can 
also be used to build a clock that reaches a finite precision with arbitrarily low dissipation.

The theoretical framework for systems driven by a protocol that changes at stochastic times considered here was crucial to obtain our main result. With 
this theory the system and external protocol together form a bipartite Markov process and quantities like the diffusion coefficient can be calculated 
with standard methods for steady states. This option represents a major advantage in relation to standard deterministic protocols that reach a periodic
steady state, where a similar method to calculate the diffusion coefficient is not available.

It is possible to consider a stochastic protocol that also has reversed jumps. In this case, the entropy production associated with generating the external protocol
is finite. This well defined quantity can then be taken into account in a way consistent with thermodynamics \cite{verl14}. 
If one chooses to also consider the entropy production due to the changes in the external protocol as part of the thermodynamic cost, then the thermodynamic uncertainty relation 
from Sec. \ref{mainsec2} is again valid. This result follows from the fact that the uncertainty relation from \cite{bara15} is valid for any Markov process,
including the full bipartite process of system and protocol together. From a physical perspective, this observation is not surprising. If we also take the cost 
of generating the stochastic protocol into account, then our full bipartite process is a thermodynamic system driven by a fixed force, which obeys the thermodynamic uncertainty relation.
For example, this cost of the external protocol would be of interest if the external protocol is driven by some chemical reaction \cite{mach16}. However, if the protocol 
is directed by some truly external process, e.g., day light changes that influence a circadian clock or an external field applied to a system, then the entropic 
cost of the external protocol is irrelevant, independent on whether the protocol is deterministic or stochastic. It is in this case that our definition 
of cost for a system driven by an external protocol is meaningful.

Finally,  the experimental confirmation of both the thermodynamic uncertainty relation for systems driven by 
a fixed thermodynamic force and the limit of high precision in the output with small dissipation for a system driven by an external periodic 
protocol remains an open challenge. Promising candidates for the experimental realization of a Brownian clock are single molecules, 
colloidal particles, and small electronic systems.

\appendix

\section{External protocols that change at stochastic times}
\label{App1}

In this appendix, we consider a theoretical framework for systems driven by periodic protocols that change at stochastic times.

\subsection{Two state model}

\begin{figure}
\includegraphics[width=87mm]{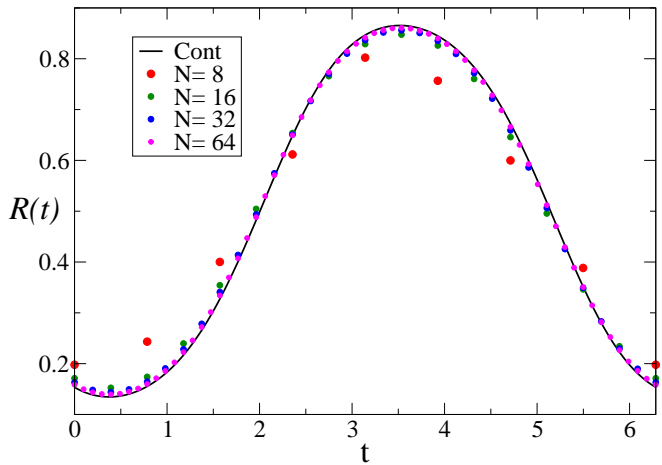}
\caption{Comparison between the periodic steady state obtained with the continuous deterministic protocol from Eq. \eqref{Et2state}
and the steady state obtained with the stochastic protocol that jumps with a rate $\gamma=L/\tau$. 
For the second case the horizontal axis is $t=nT/L$. For the periodic steady state we plot $R^{\textrm{PS}}(t)$ 
and for the steady state the conditional probability $P(u|n)$ for different values of $L$. The parameters are set to $\omega=1$ and $E_0=1$.
}
\label{fig1}
\end{figure}

As a simple example of a periodic steady state we consider a two
state system. The ``lower'' level has energy $0$ while the ``upper'' level has a time dependent
periodic energy 
\begin{equation}
E(t)=2E_{0}\cos(\omega t),\label{Et2state}
\end{equation}
where $\tau\equiv2\pi/\omega$ is the period. The transition rates fulfill
the detailed balance relation $k_{+}(t)/k_{-}(t)=\textrm{e}^{-E(t)}$.  
The master equation reads 
\begin{equation}
\frac{dR}{dt}(t)=k_{+}(t)-[k_{+}(t)+k_{-}(t)]R(t),
\end{equation}
where $R(t)$ is the probability that the level with energy $E(t)$
is occupied. With the particular choice $k_{+}=k_{-}^{-1}=\textrm{e}^{-E(t)/2}$ and 
the initial condition $R(0)=0$, the solution of this equation reads 
\begin{equation}
R(t)= \int_{0}^{t}\textrm{e}^{-E_0\cos(\omega t')}\textrm{e}^{-\int_{t'}^{t}2\cosh[E_{0}\cos(\omega t'')]dt''}dt'.\label{Pttwostate}
\end{equation}
This solution has the property that, for large $t$, the system reaches a periodic steady state independent of initial conditions
that fulfills the relation $R^{\textrm{PS}}(t)=R^{\textrm{PS}}(t+\tau)$. The function $R^{\textrm{PS}}(t)$ in a period $\tau$ obtained from 
Eq. \ref{Pttwostate} is shown in Fig. \ref{fig1}.

\begin{figure}
\includegraphics[width=87mm]{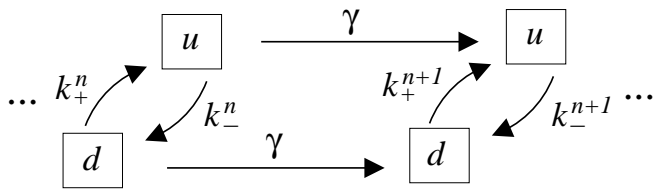}
\caption{Two state model with a stochastic protocol. The states of the system $d$ and $u$ have energy $0$ and $E^n$, respectively. 
The protocol changes from $n$ to $n+1$ with a rate $\gamma$.}
\label{fig2}
\end{figure}

Instead of an energy that changes continuously and deterministically
with time we now consider discontinuous changes that take place at
random times, as shown  in Fig. \ref{fig2}. 
Particularly, the transition rates for changes in the
state of the system are now written as $k_{\pm}^{n}$, where 
$n$ plays a role similar to $t$ in Eq. \eqref{Et2state}. The detailed
balance condition for jumps changing the state of the system reads 
$k_{+}^n/k^n_{-}=\textrm{e}^{-E^n}$. The period $\tau$ is 
partitioned in $L$ pieces, leading to $E^{n}=E(t=n\tau/L)$. 
The energy $E^{n}$ can change to $E^{n+1}$ with jumps that take place
with a rate $\gamma$, where for $n=L-1$ the jump is to $E^{n+1}=E^{0}$.
The reversed transition leading to an energy change from $E^{n+1}$ to $E^{n}$ is not allowed. 
The external protocol and the system together form a bipartite Markov process that has $2\times L$
states (see Fig. \ref{fig2}). Furthermore, the external protocol alone is a unicyclic Markov process with 
the irreversible transitions $E^{0}\to E^{1}\to\ldots\to E^{L-1}\to E^{0}$.
To match with the protocol in Eq. \eqref{Et2state}, the rate $\gamma$ is set to $\gamma=L/\tau$. 

The full Markov process of system and protocol together reaches a stationary state,
with the joint probability that the protocol is in state $n$ and the system is in a generic 
state $i$ denoted by $P_i^n$. The marginal probability of the state of the protocol is 
$P^{n}\equiv\sum_{i}P_{i}^{n}$. For the present case $P^{n}=1/L$. Comparing the periodic 
steady state with the stationary state, the quantity analogous to the probability $R^{PS}(t)$ is the conditional 
probability $P(u|n)\equiv P_{u}^{n}/P^{n}$, where $u$ denotes the state with energy $E^n$. 
This conditional probability is compared to 
$R^{PS}(t)$ in Fig. \ref{fig1}. Clearly, for larger $L$ the conditional probability of the 
steady state tends to the probability in the periodic steady state. More generally, 
in Appendix \ref{App2} we prove that for any periodic steady state it is possible to construct a 
steady state of a bipartite process with a stationary probability that converges to the probability of the periodic steady state 
in the limit $L\to\infty$.

For both protocols the system is out of equilibrium due to the time variation of the energy levels. 
For the periodic steady state the average rate of work done on the system is  
\begin{equation}
\dot{w}^{PS}\equiv \frac{1}{\tau}\int_0^\tau R^{PS}(t)\dot{E}(t)dt.
\label{workcont}
\end{equation}
The integrand is just the probability of being in the upper state with energy $E(t)$ multiplied by
the rate of energy change $\dot{E}(t)$. The expression for the rate of work done on the system for the model with 
stochastic jumps in the protocol is
\begin{equation}
\dot{w}\equiv\gamma\sum_{n}P_{u}^{n}(E^{n+1}-E^{n})=\sum_nP^n P(u|n)(E^{n+1}-E^{n}),
\label{workdisc}
\end{equation}
The sum in $n$ corresponds to the integral 
in $t$ in Eq. \eqref{workcont}, $P^n=1/L$ is the average fraction of time that the protocol spends in state $n$ during a period,  $P(u|n)$ is equivalent 
to $R^{PS}(t)$, and $E^{n+1}-E^{n}$ is related to $\dot{E}(t)$ in Eq. \eqref{workcont}. 

\begin{figure}
\centering\includegraphics[width=87mm]{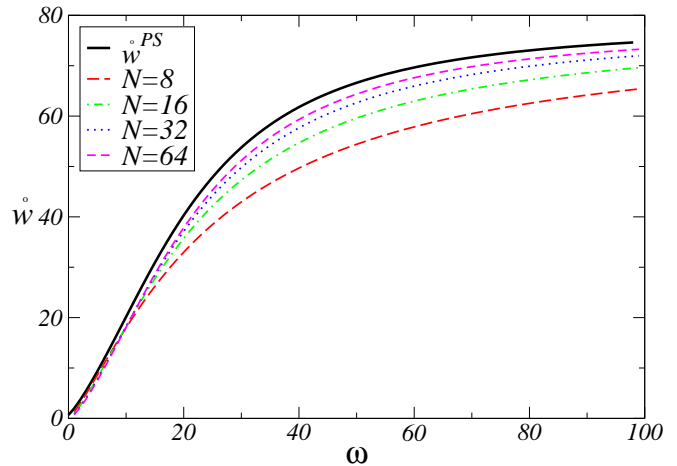}
\caption{Rate of work done by the external process $\dot{w}$ as a function of $\omega=2\pi \gamma/N$. For the periodic steady 
state this rate is denoted by $\dot{w}^{PS}$.}
\label{fig3}
\end{figure}

In Fig. \ref{fig3} we compare $\dot{w}^{PS}$ with $\dot{w}$. For large $L$, they become the same, which is a consequence of the convergence of the corresponding probabilities 
shown in Fig. \ref{fig1}. Even if for smaller $L$ the quantitative discrepancy between  $\dot{w}^{PS}$ and 
$\dot{w}$ is noticeable, the qualitative behavior is still similar, i.e., in all cases the rate of work done on the system is an 
increasing function of $\omega$. 

\subsection{General theory}

We now consider the general case that includes an arbitrary network of states beyond the ring geometry
of the models in the main text, which is similar to the framework from \cite{verl14}. The system and the external protocol together form a Markov process with 
states labeled by the variables $i=1,2,\ldots,N$ for the state of the system and $n=0,1,\ldots,L-1$
for the state of the external protocol. This full Markov process is bipartite, i.e., a transition changing 
both variables is not allowed \cite{hart14}. A state of the system $i$ with the external protocol 
in state $n$ has free energy $E_{i}^{n}$. The transition rates for a change in the
state of the system fulfill the generalized detailed balance relation \cite{seif12}
\begin{equation}
\frac{k_{ij}^{n}}{k_{ji}^{n}}=\exp\{E_{i}^{n}-E_{j}^{n}+\mathcal{A}^n d_{ij}\},
\label{gendet}
\end{equation}
where $\mathcal{A}^n$ is a thermodynamic force or affinity and $d_{ij}$ is a generalized distance. For example,
if the transition from $i$ to $j$ is related to a chemical reaction then $\mathcal{A}^n$ is the chemical potential difference driving the 
reaction and $d_{ij}$ is the number of molecules consumed in the reaction.

A jump changing the external protocol from $(i,n)$ to $(i,n+1)$ takes place with rate $\gamma_n$, while the reversed jump 
is not allowed. The master equation for the full bipartite process then reads
\begin{align}
\frac{d}{dt'}P_{i}^n(t')& =  \sum_{j}\left[P_{j}^{n}(t')k_{ji}^{n}-P_{i}^{n}(t')k_{ij}^{n}\right] \nonumber\\
& +\left[\gamma_{n-1}P_i^{n-1}(t')-\gamma_{n}P_i^{n}(t')\right],
\label{Meq}
\end{align}
where $P_i^n(t')$ is the probability that the system is at state $i$ and the external protocol at state $n$ at time $t'$. We use the variable 
$t'$ in this master equation in order to stress the difference with the variable $t$ used for the periodic steady state.
In the following we consider only the stationary distribution, which is simply denoted $P_i^n$. 

The entropy production, which characterizes the rate of dissipated heat in an isothermal system, is defined as
\begin{equation}
\sigma\equiv\sum_{n}\sum_{ij}P_{i}^{n}k_{ij}^{n}\ln\frac{k_{ij}^{n}}{k_{ij}^{n}}\ge 0.
\label{sgen}
\end{equation}
The above inequality is demonstrated in \cite{hart14}. This entropy production does not include 
jumps that lead to a change in the external protocol. The mathematical expression for the entropy
production of the full Markov process also contains a contribution that comes from these jumps. This 
contribution is related to the entropy production due to the external protocol \cite{verl14} (see also \cite{mach16}). As
usual for thermodynamic systems driven by an external protocol, we do not take such contribution, which is 
irrelevant for the second law in Eq. \eqref{sgen}, into account.

The first law reads 
\begin{equation}
\dot{w}\equiv \dot{E}+\dot{q},
\end{equation}
where $\dot{w}$ is the rate of work done on the system and $\dot{E}$ is the rate of increase of the internal energy. Since $k_BT=1$, the
rate of dissipated heat is $\dot{q}=\sigma$. In the stationary state 
\begin{equation}
\dot{E}\equiv \frac{d}{dt}\sum_{n,i}P_i^nE_i^n=0,
\end{equation}
which, with Eq. \eqref{Meq}, leads to the equation
\begin{equation}
\sum_{n,ij}(P_j^nk_{ji}^n-P_i^nk_{ij}^n)E_i^n=\sum_{n,i}(P_i^{n}\gamma_n-P_i^{n-1}\gamma_{n-1})E_i^n.
\label{eqEE}
\end{equation}
In the stationary state the first law then reads $\dot{w}= \dot{q}$. Using equation 
\eqref{eqEE} we can rewrite the entropy production \eqref{sgen} in the form 
\begin{equation}
\sigma=\sum_{n}P^n\left(\sum_{i<j}J_{ij}^{n}d_{ij}\mathcal{A}^{n}+\gamma_{n}\sum_{i}P(i|n)(E_{i}^{n+1}-E_{i}^{n})\right),
\label{sgen2}
\end{equation}
where $J_{ij}^{n}\equiv P(i|n)k_{ij}^{n}-P(j|n)k_{ji}^{n}$ is a probability current. The second term on the right hand side of this equation is 
the work done by the external variation of the protocol. The first term is the work related to the affinity $\mathcal{A}^n$; this term would be 
present even if the protocol was constant in time. For the model considered in Sec. \ref{mainsec3} of the main text only the second term is present.

We now compare expression \eqref{sgen2} with the expression for entropy production for a standard 
periodic steady state. The master equation for the periodic steady state is   
\begin{equation}
\frac{d}{dt}R_{i}= \sum_{j}[R_{j}(t)k_{ji}(t)-R_{i}(t)k_{ij}(t)].
\label{Meqcont}
\end{equation}
where $R_i(t)$ is the probability of the system being in state $i$ at time $t$.
The generalized detailed balance relation \eqref{gendet} in this case reads  
\begin{equation}
\frac{k_{ij}(t)}{k_{ji}(t)}=\exp[E_{i}(t)-E_{j}(t)+\mathcal{A}(t)d_{ij}],
\end{equation}
where the time dependent quantities have a period $\tau$. We assume that for large $t$
Eq. \eqref{Meqcont} reaches a periodic steady state with the property $R^{PS}_{i}(t)=R^{PS}_{i}(t+\tau)$.

From the average energy 
\begin{equation}
E^{PS}(t)\equiv \sum_i R_i^{PS}(t)E_i(t)
\end{equation}
that is also periodic, i.e.,
\begin{equation}
E^{PS}(\tau)-E^{PS}(0)=\int_{0}^{\tau}\dot{E}^{PS}(t)dt=0,
\end{equation}
we obtain
\begin{align}
\int_{0}^{\tau}\sum_{ij}[R_i^{PS}(t)k_{ij}(t)-R_j^{PS}(t)k_{ji}(t)]E_i(t)dt\nonumber\\
= \int_{0}^{\tau}\sum_{i} R_i^{PS}(t)\dot{E}_i(t)dt.
\end{align}
This equation is equivalent to Eq. \eqref{eqEE}.
The standard entropy production rate from stochastic thermodynamics \cite{seif12} for this periodic steady state 
is 
\begin{align}
 & \sigma^{PS}  \equiv \frac{1}{\tau}\int_0^\tau R_i^{PS}(t)k_{ij}(t)\ln\frac{k_{ij}(t)}{k_{ji}(t)}dt\nonumber\\
& =\frac{1}{\tau}\int_{0}^{\tau}\left(\sum_{i<j}J_{ij}(t)d_{ij}\mathcal{A}(t)+\sum_{i}R_i^{PS}(t)\dot{E}_i(t)\right)dt,
\end{align}
where $J_{ij}(t)\equiv R_{i}^{PS}(t)k_{ij}(t)-R_{j}^{PS}k_{ji}(t)$. This expression is analogous to the entropy production \eqref{sgen2}.

The problem of determining a periodic steady state probability analytically is typically complicated, whereas 
finding the probability distribution of a steady state in the case of stochastic changes in the external protocol can be much easier. This framework should then be useful also for 
the analysis of the qualitative behavior displayed by a system driven by a deterministic external protocol that is preserved in the case of a discretized stochastic protocol.

\subsection{Diffusion coefficient}

A main advantage of the stochastic protocols we consider here is that we can determine the diffusion coefficient defined in Eq. \eqref{eqdiff}.
For a general model defined by the master equation \eqref{Meq}, we calculate the diffusion coefficient associated with an elementary current between states 
$a$ and $b$: the random variable $X$ in Eq. \eqref{eqdiff} is such that if there is a jump from $a$ to $b$ it increases by one and if there is jump from $b$ to $a$ it decreases
by one. 

This random variable is a standard probability current of a steady state, therefore, the method from Koza \cite{koza99} (see also \cite{bara15a,bara15}) can be 
used to calculate the current and diffusion coefficient in the following way. The $N$-dimensional matrix $\mathbf{L}^n(z)$, where $z$ is a real variable, is defined as  
\begin{equation}
\mathbf{L}^n_{ji}(z)\equiv\begin{cases} k_{ij}^n\textrm{e}^{z(\delta_{i,a}\delta_{j,b}-\delta_{i,b}\delta_{j,a})}\qquad\textrm{if }i\neq j \\ -(\sum_{l}k_{il}^n+\gamma_n)\qquad\textrm{if }i=j\end{cases}.
\label{Lmat}
\end{equation}
The modified generator \cite{koza99,lebo99} associated with the current $X$ is a matrix with dimension $N\times L$ given by
\begin{equation}
\mathbf{L}(z)\equiv\left(\begin{array}{cccc}
\mathcal{\mathbf{L}}_{0}(z)-\mathbf{\Gamma}_0 & 0 & \ldots & \mathbf{\Gamma}_{L-1}\\
 \mathbf{\Gamma}_0 & \mathcal{\mathbf{L}}_{1}(z)-\mathbf{\Gamma}_1 &   \ldots  & 0\\
 0 & \mathbf{\Gamma}_1 &   \ldots & 0 \\
\vdots & \vdots &  \ddots  & \vdots\\
0 & 0 &  \ldots & \mathcal{\mathbf{L}}_{L-1}(z)-\mathbf{\Gamma}_{L-1}
\end{array}\right),\label{statfull}
\end{equation}
where $\mathbf{\Gamma}_n$ is the identity matrix with dimension $N$ multiplied by $\gamma_n$. As explained in \cite{bara15a,bara15}, we can obtain
$J$ and $D$, defined in Eqs. \eqref{eqcurr} and \eqref{eqdiff}, respectively,  from the coefficients $C_m(z)$ of the characteristic polynomial associated with $\mathbf{L}(z)$, which are  defined through the relation 
\begin{equation}
\sum_{m=0}^{NL} C_m(z)x^m\equiv\textrm{det}[x\mathbf{I}-\mathbf{L}(z)]. 
\end{equation}
The current and diffusion coefficient are given by \cite{koza99}
\begin{equation}
J=-C_0'/C_1
\label{currco}
\end{equation}
and
\begin{equation}
D=-(C_0''+2C_1'J+2C_2J^2)/2C_1,
\label{diffco}
\end{equation}
where the lack of dependence in $z$ indicates evaluation of the function at $z=0$ and the primes denote derivatives with respect to $z$.

\section{Proof of the equivalence between periodic steady state and steady state of a bipartite process}
\label{App2}

In this appendix we prove that for any given periodic steady state it is possible to construct a bipartite process
that has a stationary distribution corresponding to the distribution of the periodic steady state. 

We consider a periodic steady state following the master equation \eqref{Meqcont}, which can be written in the form
\begin{equation}
\frac{d\mathbf{R}(t)}{dt}=\mathcal{\mathbf{M}}(t)\mathbf{R}(t),
\label{eqperi}
\end{equation}
where stochastic matrix $\mathcal{\mathbf{M}}(t)$ has period $\tau$, i.e., 
$\mathcal{\mathbf{M}}(t)= \mathcal{\mathbf{M}}(t+\tau)$, and $\mathbf{R}(t)$ is the probability vector with $N$ states. 
The  periodic steady state $\mathbf{R}^{PS}(t)$.

The period $\tau$ is discretized in $L$ small intervals so that in each time interval the transition
rates can be taken as time-independent. In the nth-time interval the system then follows the master 
equation with time independent transition rates 
\begin{equation}
\frac{d\mathbf{R}_{n}}{dt}=\mathcal{\mathbf{M}}_{n}\mathbf{R}_{n},
\end{equation}
where $\mathcal{\mathbf{M}}_{n}\equiv \mathcal{\mathbf{M}}(n\tau/L)$ and $\mathbf{R}_{n}\equiv \mathbf{R}^{PS}(n\tau/L)$. 
The formal solution of this equation is 
\begin{equation}
\mathbf{R}_{n}^{(f)}=\mathbf{\exp(\mathbf{M}_{n}\epsilon)R}_{n}^{(i)},\label{soldisc}
\end{equation}
where $\epsilon\equiv\tau/L$ and the superscript $i$ ($f$) denotes the
initial (final) distribution of the system in the time interval $[n\tau/L,(n+1)\tau/L]$.
Using the relation $\mathbf{R}_{n}^{(f)}=\mathbf{R}_{n+1}^{(i)}$ we rewrite Eq. (\ref{soldisc}) for $n+1$ as  
\begin{equation}
\mathbf{\exp(-\mathbf{M}_{n+1}\epsilon)R}_{n+1}^{(f)}=\mathbf{R}_{n}^{(f)},
\end{equation}
where we have multiplied the equation by $\mathbf{\exp(-\mathbf{M}_{n+1}\epsilon)}$. 
Expanding to first order in $\epsilon$ we obtain
\begin{equation}
\mathbf{R}_{n}^{(f)}=(1-\mathcal{\mathbf{M}}_{n+1}\epsilon)\mathbf{R}_{n+1}^{(f)}\equiv\tilde{\mathbf{M}}{}_{n+1}\mathbf{R}_{n+1}^{(f)},\label{solddisc2}
\end{equation}
which is valid for $n=0,1,\ldots,L-1$. Due to the periodicity for
$n=L-1$ this equation reads $\mathbf{R}_{L-1}^{(f)}=\tilde{\mathbf{M}}{}_{0}\mathbf{R}_{0}^{(f)}$. Therefore, Eq. (\ref{solddisc2}) leads to  
\begin{equation}
\mathbf{R}_{n}^{(f)}=\tilde{\mathbf{M}}{}_{n+1}\tilde{\mathbf{M}}{}_{n+2}\ldots\tilde{\mathbf{M}}{}_{L-1}\tilde{\mathbf{M}}{}_{0}\ldots\tilde{\mathbf{M}}{}_{n}\mathbf{R}_{n}^{(f)},
\label{soldisc3}
\end{equation}
i.e., $\mathbf{R}_{n}^{(f)}$ is the eigenvector of the matrix $\tilde{\mathbf{M}}{}_{n+1}\tilde{\mathbf{M}}{}_{n+2}\ldots\tilde{\mathbf{M}}{}_{L-1}\tilde{\mathbf{M}}{}_{0}\ldots\tilde{\mathbf{M}}{}_{n}$ associated with
the eigenvalue $1$. 

We now construct a bipartite process with a steady state corresponding to the periodic steady state
$\mathbf{R}^{PS}(t)$. The Markov process including both the system and the
external protocol has $N\times L$ states, which is the dimension of 
the stationary distribution  vector $\mathbf{P}$. The stochastic matrix that fulfill the relation
$\mathbf{L}\mathbf{P}=0$ can be written in the form
\begin{equation}
\mathbf{L}=\left(\begin{array}{ccccc}
\mathcal{\mathbf{L}}_{0}-\mathbf{\Gamma} & 0 & \ldots & \mathbf{\Gamma}\\
\mathbf{\Gamma} & \mathcal{\mathbf{L}}_{1}-\mathbf{\Gamma} &  \ldots  & 0\\
0 & \mathbf{\Gamma} &  \ldots  & 0\\
\vdots & \vdots & \ddots & \vdots \\
0 & 0 & \ldots &  \mathcal{\mathbf{L}}_{L-1}-\mathbf{\Gamma}
\end{array}\right),\label{statfull2}
\end{equation}
where $\mathbf{\Gamma}$ is the identity matrix with dimension $N$ multiplied
by $\gamma$, and $\mathcal{\mathbf{L}}_{n}$ is the matrix in Eq. \eqref{Lmat} with $z=0$ and $\gamma_n=\gamma$. 
From (\ref{statfull}), the stationary master equation can
be written as 
\begin{equation}
\mathbf{L}_{n+1}\mathbf{P}_{n+1}+\gamma\mathbf{P}_{n}-\gamma\mathbf{P}_{n+1}=0,
\label{Lrec}
\end{equation}
where $\mathbf{P}_{n}$ is a vector that contains the $N$ states
of the system for the protocol in state $n$. This equation is valid
for $n=0,1,\ldots,L-1$, where if $n=L-1$ then $n+1=0$. Eq. (\ref{Lrec}) implies
\begin{equation}
\mathbf{P}_{n}=\tilde{\mathbf{L}}{}_{n+1}\tilde{\mathbf{L}}{}_{n+2}\ldots\tilde{\mathbf{L}}{}_{L-1}\tilde{\mathbf{L}}{}_{0}\ldots\tilde{\mathbf{L}}{}_{n}\mathbf{P}_{n},
\label{solddisc4}
\end{equation}
where $\tilde{\mathbf{L}}{}_{n}\equiv\mathbf{1}-\mathbf{L}_{n}\gamma^{-1}$. Hence, $\mathbf{P}_{n}$ is the eigenvector of 
$\tilde{\mathbf{L}}{}_{n+1}\tilde{\mathbf{L}}{}_{n+2}\ldots\tilde{\mathbf{L}}{}_{N-1}\tilde{\mathbf{L}}{}_{0}\ldots\tilde{\mathbf{L}}{}_{n}$
associated with the eigenvalue $1$. Comparing (\ref{soldisc3}) with (\ref{solddisc4}), we obtain that the choices 
$\mathbf{L}_{n}=\mathbf{M}_{n}$ and $\gamma=\epsilon^{-1}=L/\tau$ lead to $\mathbf{P}_{n}\propto \mathbf{R}_{n}^{(f)}$. These two quantities are not 
exactly the same due to a different normalization, i.e., $\sum_i P_i^n=1/L$. Therefore, the steady state of the stochastic matrix \eqref{statfull} in the limit $L\to\infty$,
with $\gamma=L/\tau$ and $\mathbf{L}_{n}=\mathbf{M}(n\tau/L)$, is equivalent to the periodic steady state from Eq. \eqref{eqperi}.

\section{Details for the model from Sec. \ref{mainsec3}}
\label{App3}

In this Appendix we define more precisely the model from Sec. \ref{mainsec3} with changes in the energies and energy barriers that take place at random times, 
and explain how we calculate $J$, $D$, and $\sigma$. 

The clock and external protocol together form a bipartite Markov process. The model is defined by the stochastic matrix for this 
bipartite process. This matrix is of the form \eqref{statfull} with 
\begin{align}
& (\mathbf{L}_n)_{i+1i}=\chi_{i-n}\epsilon_{i-n},\nonumber\\
& (\mathbf{L}_n)_{i-1i}=\chi_{i-1-n}\epsilon_{i-n},\nonumber\\
& (\mathbf{L}_n)_{ii}=-(\chi_{i-n}+\chi_{i-1-n})\epsilon_{i-n},
\end{align}
where the other elements of the matrix are $0$. For this model the number of jumps that change the protocol is $L=N$.

Due to the symmetry of the external protocol, the fluctuating current between states $N$ and $1$, which we label $X$, is the same as 
the fluctuating current between any pair of states $i$ and $i+1$. The random variable $X$ is then the sum of all these currents divided by $N$. 
The statistics of this random variable can be described by a matrix that has dimension $N$ instead 
of the full matrix for the bipartite process that has dimension $N^2$.  This reduction can be demonstrated in the following way. Instead of changing the
transition rates between states after a jump with rate $\gamma$ we consider that the states rotate in the anti-clock wise direction. In this case a label $\alpha=1$
refers to the states that have transition rate $\epsilon_1\chi_1$ to jump to state $\alpha=2$ and transition rate $\epsilon_1\chi_N$  to jump to state $\alpha=N$. This 
label $\alpha$ that marks the state that has certain transition rates is different from the label $i$ that marks a position in the ring. The sum 
of the currents  between the states with the labels $i$ is the same as the sum of currents between states with label $\alpha$.  Within the 
label $\alpha$ a jump with rate $\gamma$, which is related to a change in the external protocol, implies a jump from $\alpha$ to $\alpha-1$. 
Therefore, instead of a stochastic matrix of the form \eqref{statfull} the time evolution of the probability vector of the states $\alpha=1,2,\ldots,N$ is 
described by the stochastic  matrix $\mathbf{L}^*$ that is defined by the following non-zero elements,
\begin{align}
& \mathbf{L}^*_{\alpha+1\alpha}=\chi_{\alpha}\epsilon_{\alpha},\nonumber\\
& \mathbf{L}^*_{\alpha-1\alpha}=\chi_{\alpha-1}\epsilon_{\alpha}+\gamma,\nonumber\\
& \mathbf{L}^*_{\alpha\alpha}=-(\chi_{\alpha}+\chi_{\alpha-1})\epsilon_{\alpha}-\gamma.
\end{align}
With this reduction the system and protocol together are described by a matrix with dimension $N$. The modified generator \eqref{statfull}
is also reduced to a $N$-dimensional matrix $\mathbf{L}^*(z)$. Its non-zero elements are 
\begin{align}
& \mathbf{L}^*(z)_{\alpha+1\alpha}=\chi_{\alpha}\epsilon_{\alpha}\textrm{e}^{z/N},\nonumber\\
& \mathbf{L}^*(z)_{\alpha-1\alpha}=\chi_{\alpha-1}\epsilon_{\alpha}\textrm{e}^{-z/N}+\gamma,\nonumber\\
& \mathbf{L}^*(z)_{\alpha\alpha}=-(\chi_{\alpha}+\chi_{\alpha-1})\epsilon_{\alpha}-\gamma.
\end{align}
The current $J$ and the diffusion coefficient $D$ are given by relations \eqref{currco} and \eqref{diffco}, respectively, with the coefficients 
$C_m(z)$ given by
\begin{equation}
\sum_{m=0}^{N} C_m(z)x^m\equiv\textrm{det}[x\mathbf{I}-\mathbf{L}^*(z)]. 
\end{equation}
The entropy production $\sigma$ is calculated with relation \eqref{sgen2}.

We now consider the model in the limit $\chi_N=0$, $\chi_1=\chi_2=\ldots=\chi_{N-1}=\chi$, and $\chi\gg \gamma$.
The condition $\chi\gg \gamma$ means that the system reaches an equilibrium distribution $P^*_\alpha$ before a jump 
with rate $\gamma$ takes place. This equilibrium distribution is given by 
\begin{equation}
P^*_\alpha= \textrm{e}^{-E_\alpha}/Z,
\end{equation}
where $Z=\sum_{\alpha=1}^N\textrm{e}^{-E_\alpha}$. With this distribution we can calculate the entropy production rate $\sigma$ given in Eq. \eqref{Slim} using Eq. \eqref{sgen2}.

The total current $X$ is the sum of the current between all states divided by $N$. Denoting the current 
between $\alpha$ and $\alpha+1$ by $X_{\alpha\alpha+1}$ we obtain $X=(X_{12}+X_{23}+\ldots+X_{1N})/N$. 
The fluctuating current through the links associated with the rate $\gamma$ that leave state $\alpha$ is denoted by $Y_\alpha$. The average value 
for this unidirectional current is $\gamma P^*_\alpha$. From Kirchhoff's law for the fluctuating currents we obtain 
\begin{equation}
X=\sum_{\alpha=2}^N Y_\alpha/N-(N-1)Y_1/N.
\end{equation}
Hence, the random variable $X$ can be viewed as a biased random walk that gives a step of size $1/N$ forward if the protocol changes and the clock is 
in a state $\alpha\neq 1$ or a step of size $(N-1)/N$ backward if the clock is in state $\alpha=1$. The master equation for this random walk reads 
\begin{align}
\frac{d}{dt}P(X,t) &= \kkp P(X-1/N,t)+\kkm P(X+1-1/N,t)\nonumber\\
&-(\kkp+\kkm)P(X,t),
\end{align}
where $\kkp\equiv\gamma\sum_{\alpha=2}^N P^{*}_\alpha$ and $\kkm\equiv\gamma P^{*}_1$. Using the Laplace transform 
\begin{equation}
\tilde{P}(z,t)\equiv\sum_X P(X,t)\textrm{e}^{Xz}
\end{equation}
we obtain
\begin{align}
\frac{d}{dt}\tilde{P}(z,t) &= \bigg[\kkp \textrm{e}^{z/N}+\kkm\textrm{e}^{-(N-1)z/N}\nonumber\\
&-(\kkp+\kkm)\bigg]\tilde{P}(z,t).
\end{align}
The solution of this differential equation with boundary condition $\tilde{P}(0,t)=1$ is $\tilde{P}(z,t)= \textrm{e}^{\psi(z)t}$, with
\begin{equation}
\psi(z)= \kkp\textrm{e}^{z}+\kkm\textrm{e}^{-(N-1)z}-(\kkp+\kkm).
\end{equation}
From this solution we obtain
\begin{equation}
J=\left.\frac{d}{dz}\psi(z)\right|_{z=0}= [\kkp-\kkm(N-1)]/N 
\end{equation}
and
\begin{equation}
2D=\left.\frac{d^2}{dz^2}\psi(z)\right|_{z=0}= [\kkp+\kkm(N-1)^2]/N^2, 
\end{equation}
which are the expressions given in Eqs. \eqref{Jlim} and \eqref{Dlim} of the main text, respectively.

%


\begin{thebibliography}{67}
\bibitem{foot} In this case, this random variable is indeed equivalent to our random variable $X$ \cite{schn95}. This cycle 
completion time is more similar to the random variable analyzed in
\cite{cao15}, where the dispersion of the time at which a peak of a biochemical oscillation 
occurs was considered. Since, strictly speaking, for  $k_N^-\neq0$ the cycle completion 
time is,  due to the possibility of backward jumps, not a meaningful variable anymore, we 
stick with our identification of counting time.
\bibitem{foot2} This mapping of a periodic steady state onto a bipartite system with a larger number of states is different from the mapping introduced in \cite{raz16}, where
systems driven by a fixed force are mapped onto systems driven by an external protocol, and {\sl vice versa}, with the same number of states.
\end{thebibliography}

 \end{document}